\let\oldvec\vec
\documentclass{llncs}
\let\vec\oldvec

\usepackage[utf8]{inputenc}
\usepackage[T1]{fontenc}
\usepackage{amsmath, amssymb}
\usepackage{mathtools}
\usepackage{hyperref}
\usepackage{tikz}
\usepackage{pgfplots}
\usepackage{currfile}
\newcommand{\currfileabspath}{}
\usepackage{subcaption}
\usepackage{xspace}
\usepackage{booktabs}
\usepackage{dcolumn}

\usepackage{aicescover}

\colorlet{graybg}{gray!10}
\colorlet{plot1}{red!75!white}
\colorlet{plot2}{green!75!black}
\colorlet{plot3}{blue}
\colorlet{plot4}{yellow!75!black}
\colorlet{plot5}{violet}
\colorlet{plot6}{cyan}
\colorlet{plot7}{orange}

\usetikzlibrary{
    external,
}

\pgfkeys{
    /tikz/external/mode=list and make
}
\tikzsetexternalprefix{figures/externalized/}
\tikzset{external/export=false}

\pgfplotsset{
    every axis/.append style={
        axis background/.style={fill=graybg},
        legend cell align=left,
        xlabel near ticks,
        ylabel near ticks,
        enlarge x limits={value=0.05, auto},
        enlarge y limits={value=0.05, auto},
        width=\textwidth,
        ymajorgrids=true,
    },
    every axis plot/.append style={
        thick
    },
    kernelplot/.style={
        xlabel={kernel invocation},
        ylabel={relative error [\%]},
        xmin=0,
        enlarge x limits={value=0, auto},
        every axis plot/.append style={
            only marks,
            mark size=1.5pt,
            thin
        },
    },
    dgeqr2/.style={color=plot2, mark=x},
    dlarft/.style={color=plot5, mark=x},
    dcopy/.style={ color=plot4, mark=*, mark size=.2pt},
    dtrmm/.style={ color=plot3},
    dgemm/.style={ color=plot1},
    dtrmmRLNU/.style={dtrmm, mark=x},
    dgemmTN/.style={  dgemm, mark=x},
    dtrmmRUNN/.style={dtrmm, mark=o},
    dgemmNT/.style={  dgemm, mark=o},
    dtrmmRLTU/.style={dtrmm, mark=diamond},
}

\hypersetup{
    colorlinks=true,       
    linkcolor=blue!50!black,          
    citecolor=green!50!black,        
    filecolor={.5 0 0},      
    urlcolor=red!50!black,           
}

\newcommand{\matL}[2]{\tikz[baseline=(M.base)] {
    \filldraw[fill=white, draw=gray] (0, 0) -- (0, -#1) -- (#1, -#1) -- (0, 0) node[outer sep=0, inner sep=0, midway, anchor=center] (M) {\vphantom{fg}$#2$};}
}
\newcommand{\matU}[2]{\tikz[baseline=(M.base)] {
    \filldraw[fill=white, draw=gray] (0, 0) -- (#1, 0) -- (#1, -#1) -- (0, 0) node[outer sep=0, inner sep=0, midway, anchor=center] (M) {\vphantom{fg}$#2$};}
}

\newcommand\addkernelplots[1]{
    \addplot[dgeqr2]     file {figures/data/qr/#1.dgeqr2_.dat};
    \tikzpicturedependsonfile{figures/data/qr/#1.dgeqr2_.dat}
    \label{fig:#1.dgeqr2}
    \addplot[dlarft]     file {figures/data/qr/#1.dlarft_.dat};
    \tikzpicturedependsonfile{figures/data/qr/#1.dlarft_.dat}
    \label{fig:#1.dlarft}
    \addplot[dcopy]      file {figures/data/qr/#1.dcopy_.dat};
    \tikzpicturedependsonfile{figures/data/qr/#1.dcopy_.dat}
    \label{fig:#1.dcopy}
    \addplot[dtrmmRLNU]  file {figures/data/qr/#1.dtrmm_RLNU.dat};
    \tikzpicturedependsonfile{figures/data/qr/#1.dtrmm_RLNU.dat}
    \label{fig:#1.dtrmmRLNU}
    \addplot[dgemmTN]    file {figures/data/qr/#1.dgemm_TN.dat};
    \tikzpicturedependsonfile{figures/data/qr/#1.dgemm_TN.dat}
    \label{fig:#1.dgemmTN}
    \addplot[dtrmmRUNN]  file {figures/data/qr/#1.dtrmm_RUNN.dat};
    \tikzpicturedependsonfile{figures/data/qr/#1.dtrmm_RUNN.dat}
    \label{fig:#1.dtrmmRUNN}
    \addplot[dgemmNT]    file {figures/data/qr/#1.dgemm_NT.dat};
    \tikzpicturedependsonfile{figures/data/qr/#1.dgemm_NT.dat}
    \label{fig:#1.dgemmNT}
    \addplot[dtrmmRLTU]  file {figures/data/qr/#1.dtrmm_RLTU.dat};
    \tikzpicturedependsonfile{figures/data/qr/#1.dtrmm_RLTU.dat}
    \label{fig:#1.dtrmmRLTU}
}

\newcommand\addkernelplotpic[2][]{
    \tikzpicturedependsonfile{\currfileabspath}
    \tikzsetnextfilename{#2}
    \begin{tikzpicture}
        \begin{axis}[
                kernelplot,
                yticklabel={\pgfmathparse{100*\tick}\pgfmathprintnumber{\pgfmathresult}},
                #1
            ]
            \addkernelplots{#2}
        \end{axis}
    \end{tikzpicture}
}

\newcommand\dgeqrf{{\tt dgeqrf}\xspace}
\newcommand\dtrtri{{\tt dtrtri$_\texttt{LN}$}\xspace}
\newcommand\dpotrf{{\tt dpotrf$_\texttt{U}$}\xspace}
\newcommand\dgeqr{{\tt dgeqr2}\xspace}
\newcommand\dlarft{{\tt dlarft}\xspace}
\newcommand\dcopy{{\tt dcopy}\xspace}
\newcommand\dtrmm{{\tt dtrmm}\xspace}
\newcommand\dgemm{{\tt dgemm}\xspace}
\newcommand\dtrmmRLNU{\dtrmm{}$_\texttt{RLNU}$\xspace}
\newcommand\dgemmTN{\dgemm{}$_\texttt{TN}$\xspace}
\newcommand\dtrmmRUNN{\dtrmm{}$_\texttt{RUNN}$\xspace}
\newcommand\dgemmNT{\dgemm{}$_\texttt{NT}$\xspace}
\newcommand\dtrmmRLTU{\dtrmm{}$_\texttt{RLTU}$\xspace}

\newcommand\refdgeqr{\dgeqr~(\ref*{leg:dgeqr2})\xspace}
\newcommand\refdlarft{\dlarft~(\ref*{leg:dlarft})\xspace}
\newcommand\refdcopy{\dcopy~(\ref*{leg:dcopy})\xspace}
\newcommand\refdtrmmRLNU{\dtrmmRLNU~(\ref*{leg:dtrmmRLNU})\xspace}
\newcommand\refdgemmTN{\dgemmTN~(\ref*{leg:dgemmTN})\xspace}
\newcommand\refdtrmmRUNN{\dtrmmRUNN~(\ref*{leg:dtrmmRUNN})\xspace}
\newcommand\refdgemmNT{\dgemmNT~(\ref*{leg:dgemmNT})\xspace}
\newcommand\refdtrmmRLTU{\dtrmmRLTU~(\ref*{leg:dtrmmRLTU})\xspace}

\title{
    A Study on the Influence of Caching:\texorpdfstring{\\}{}
    Sequences of Dense Linear Algebra Kernels
}
\author{Elmar Peise and Paolo Bientinesi}
\institute{
    AICES, RWTH Aachen, Germany\\
    \email{\{peise,pauldj\}@aices.rwth-aachen.de}
}

\begin{document}
    \aicescoverpage
    \maketitle

    \begin{abstract}
    It is universally known that caching is critical to attain high-performance
    implementations: In many situations, data locality (in space and time)
    plays a bigger role than optimizing the (number of) arithmetic floating
    point operations. In this paper, we show evidence that at least for linear
    algebra algorithms, caching is also a crucial factor for accurate
    performance modeling and performance prediction.
\end{abstract}


    \section{Introduction}
    \label{sec:intro}
    In dense linear algebra (DLA), very basic yet highly tuned kernels --- such as
the Basic Linear Algebra Subprograms (BLAS) --- are used as building blocks for
high level algorithms --- such as those included in the Linear Algebra PACKage
(LAPACK).  The objective of our research is to develop performance models for
those building blocks, aiming at predicting the performance of high level
algorithms avoiding entirely to execute them.  In a recent
article~\cite{modeling1}, we introduced a methodology for modeling and
predicting performance, and showed its effectiveness in ranking different
algorithmic variants solving the same target operation.  However, to accurately
tune algorithmic parameters such as the block-size, predictions of significantly
higher precision are required.  Intuitively, one would attempt to resolve this
issue through performance models of higher accuracy.  Unfortunately, beyond a
certain level higher accuracy in the models of the building blocks does not
translate into more precise predictions. In this paper we illustrate that such a
mismatch is due to the influence of CPU caching on the performance of the
compute kernels.

Several other works on the influence of caching on DLA performance exist; some
notable examples are given in the following.  Whaley empirically tunes the
block-size for LAPACK routines and emphasizes its impact on
performance~\cite{lapacktuning}.  Lam~et~al. study caching in the context of
blocking within DLA kernels~\cite{blocking1991}.  Iakymchuk~et~al. model the
number of cache misses analytically based on a very detailed analysis of kernel
implementations~\cite{roman}.

The rest of this paper is structured as follows.  We introduce the considered
problem and setup in \autoref{sec:problem} and establish bounds for the kernel
execution times in \autoref{sec:timings}.  Then, we develop a cache prediction
model in \autoref{sec:cache} and apply it to a broader range of scenarios in
\autoref{sec:results}.


    \section{The Problem}
    \label{sec:problem}
    In order to better understand the influence of caching on the performance of
compute kernels, we focus on a specific, yet exemplary algorithm and setup:  On
one core of a quadcore {\sc Intel Harpertown E5450}, we analyze the performance of
LAPACK's QR decomposition (\dgeqrf) linked to {\sc
OpenBLAS}~v.~0.2.8~\cite{openblas} on a square matrix of size\footnote{
    With $n = 1{,}568 = 2^5 \cdot 7^2$, we choose a matrix size that is not a
    power of $2$ to avoid problem size specific performance artifacts.
} $n = 1{,}568$.  With a size of 18 MB, this matrix
exceeds this CPU's largest cache (L2), consisting of 6 MBs per 2 cores.

The routine \dgeqrf implements a blocked algorithm and traverses the input
matrix from the top left to the bottom right corner, in steps of a prescribed
block-size $b$.  We fix this block-size --- this routine's only optimization
parameter --- at $b = 32$.  Within each step of the blocked traversal, \dgeqrf
executes the following sequence of kernels: \dgeqr (unblocked QR), \dlarft (form
triangular factor $T$ for the compact representation of $Q$), $b$ \dcopy{}s
(together transpose a matrix panel), \dtrmmRLNU (triangular matrix-matrix
product)\footnote{
    The subscripts {\tt R} through {\tt U} are the values of the flag arguments
    {\tt side}, {\tt uplo}, {\tt trans}, and {\tt diag}; they distinguish the
    form of the operation performed by the kernel.
}, \dgemmTN (matrix-matrix product), \dtrmmRUNN, \dgemmNT, and \dtrmmRLTU.

\begin{figure}[t]
    \centering
    \ref*{fig:routinelegend}

    \vspace{1ex}

    \tikzset{external/export=true}
    \begin{subfigure}[t]{.49\textwidth}
        \centering
        \tikzpicturedependsonfile{\currfileabspath}
        \tikzsetnextfilename{instr}
        \begin{tikzpicture}
            \begin{axis}[
                kernelplot,
                ylabel={time [cycles]},
                ymin=0,
                legend columns=-1,
                legend entries={
                    {\dgeqr\ \ },
                    {\dlarft\ \ },
                    {\dcopy\ \ },
                    {\dtrmmRLNU\ \ },
                    {\dgemmTN\ \ },
                    {\dtrmmRUNN\ \ },
                    {\dgemmNT\ \ },
                    {\dtrmmRLTU}
                },
                legend to name=fig:routinelegend,
            ]
                \addlegendimage{dgeqr2,    thin, mark size=2pt};
                \label{leg:dgeqr2}
                \addlegendimage{dlarft,    thin, mark size=2pt};
                \label{leg:dlarft}
                \addlegendimage{dcopy,     thin, mark size=1pt};
                \label{leg:dcopy}
                \addlegendimage{dtrmmRLNU, thin, mark size=2pt};
                \label{leg:dtrmmRLNU}
                \addlegendimage{dgemmTN,   thin, mark size=2pt};
                \label{leg:dgemmTN}
                \addlegendimage{dtrmmRUNN, thin, mark size=2pt};
                \label{leg:dtrmmRUNN}
                \addlegendimage{dgemmNT,   thin, mark size=2pt};
                \label{leg:dgemmNT}
                \addlegendimage{dtrmmRLTU, thin, mark size=2pt};
                \label{leg:dtrmmRLTU}
                \addkernelplots{instr}
            \end{axis}
        \end{tikzpicture}
        \caption[]{in-algorithm}
        \label{fig:instr}
    \end{subfigure}
    \begin{subfigure}[t]{.49\textwidth}
        \centering
        \addkernelplotpic[ymin=-.35, ymax=.05]{repeat-i}
        \caption{
            repeated execution
        }
        \label{fig:i-repeat}
    \end{subfigure}

    \caption[]{
        In-algorithm timings and comparison with repeated
        execution.  Along the $x$-axis, we enumerate the $1{,}873$ kernel
        invocations within \dgeqrf.
    }
    \tikzset{external/export=false}
\end{figure}
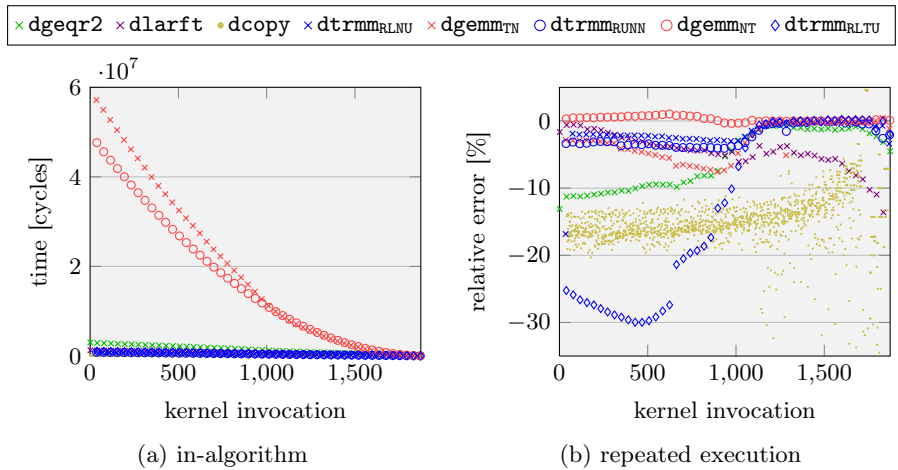

To measure the execution time of the kernels within \dgeqrf (henceforth called
{\em in-algorithm timings}), we manually instrument this routine, and collect
timestamps\footnote{
    Read from the CPU's time stamp counter through the assembly instruction
    {\tt rdtsc}.
} between kernel invocations.  The in-algorithm timings computed from these
timestamps are presented in \autoref{fig:instr}:  Along the $x$-axis, we
enumerate the $1{,}873$ kernel invocations; along the $y$-axis we present
timings of each invocation grouped by the type of kernels.  The figure shows
that the execution time is dominated by the two \dgemm kernels
(\ref{leg:dgemmTN} and \ref{leg:dgemmNT}); notably, although the size of their
operands is the same, the corresponding timings differ significantly.  Our
ultimate goal is to develop performance models that accurately predict such
differences and all other features of the in-algorithm timings.

To focus on the cache related performance features, we here attempt to
reconstruct the in-algorithm timings with a very elementary timing setup:
repeated execution of the kernels independent from each other.  In these
executions, we use the same flags and matrix sizes as those used within \dgeqrf,
and for each  operand we use a well separated memory location.  The relative
error in execution time of the median of 100 such independent repetitions
compared to the in-algorithm timings is shown in \autoref{fig:i-repeat}.  While
the relative error for \refdcopy is rather large, the total contribution of the
$1,536$ \dcopy{}s to the total runtime is below $1\%$.  Not considering these
\dcopy{}s, the absolute errors of the instrumented timings relative to the
in-algorithm timings averaged across kernel invocations (in the following simply
referred to as error) is $4.48\%$.

For most routines and especially for \refdtrmmRLTU and \refdgeqr, the repeated
execution underestimates the in-algorithm timings for the first $1{,}000$ kernel
invocations.  More surprisingly however, \dgemmNT is even overestimated --- it
is faster within \dgeqrf.


    \section{Cache-Aware Timings}
    \label{sec:timings}
    The change in behavior noticeable around the $1{,}000$th kernel invocation (see
\autoref{fig:i-repeat})
is directly linked to the size of the cache.  While traversing the matrix,
\dgeqrf only operates on the bottom right quadrant, which becomes smaller at
each iteration.  Beyond the $1{,}000$th invocation, the quadrant is small enough
to fit in the L2 cache.  As a result,  the subsequent runtime measurements
of repeated executions show only minimal differences with respect to the
in-algorithm timings.  This confirms the cache as the cause of the
discrepancies.

To better understand the scope of this influence we now manipulate the cache
locality of the kernel's operands in our independent executions.  To do so, we
assume a simplified cache replacement policy: a fully associative Least Recently
Used (LRU) algorithm.  We consider the two extreme scenarios in which the
operands immediately required by the kernels are either entirely within the L2
cache or not at all.  These in- and out-of-cache scenarios serve, respectively,
as lower and upper bounds on the in-algorithm timings.

\begin{figure}[t]
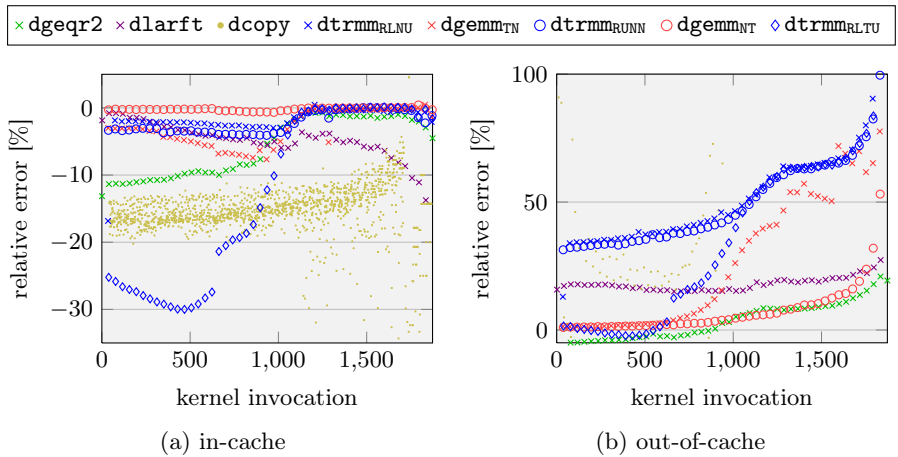

    \centering
    \ref*{fig:routinelegend}

    \vspace{1ex}
    \tikzset{external/export=true}

    \begin{subfigure}[b]{.49\textwidth}
        \centering
        \addkernelplotpic[ymin=-.35, ymax=.05]{ic-i}
        \caption{in-cache}
        \label{fig:i-ic}
    \end{subfigure}
    \begin{subfigure}[b]{.49\textwidth}
        \centering
        \addkernelplotpic[ymin=-.05, ymax=1]{ooc-i}
        \caption{out-of-cache}
        \label{fig:i-ooc}
    \end{subfigure}

    \tikzset{external/export=false}
    \caption[]{
        In-cache and out-of-cache compared to in-algorithm timings.
        In (b), the error for \refdcopy is around $1{,}000\%$.
    }
    \label{fig:i-icooc}
\end{figure}

For kernels with operands whose size is smaller than 6MBs, repeated execution
suffices to guarantee that the operands are in cache prior to execution.  By
contrast, when the aggregate size of all kernel operands exceeds 6MB (as for
\refdgemmNT), different kernel implementations (i.e. different libraries) may
initially access different regions of the operands.  An ideal in-cache setup
would place exactly the immediately accessed regions in cache.  However, since
we do not assume knowledge about kernel implementation, we restrict our in-cache
setup to fulfill the reasonable assumption that input operands are accessed
before (input-)output and output operands.  In order to accordingly prepare
the cache, we touch\footnote{
    By touching, we mean a simple read+write access to the data, e.g. $x
    \coloneqq x + \varepsilon$.
} all input operands just before the kernel invocation.  This timing setup
yields the runtime predictions shown in \autoref{fig:i-ic}.  Here, the
predictions are in all cases equal to or underestimating the in-algorithm
timings.  The error is $4.51\%$.

Under the assumption of a fully associative LRU cache, to ensure that the
operands are not in the cache, it suffices to  touch a section of the main
memory larger than the cache size.  This approach yields the runtime predictions
presented in \autoref{fig:i-ooc}.  Now, almost all predictions are equal to or
overestimating the in-algorithm timings.  The error is $29.1\%$.

Not only do the established in-cache and out-of-cache timings indeed serve as
lower and upper bounds on the in-algorithm timings, for most kernel invocations
one of these two bounds is actually attained (see \autoref{fig:i-icooc}).  Based
on this observation, the next section introduces a cache model to use these
in-core and out-of-core timings to estimate the in-algorithm timings.


    \section{Modeling the Cache}
    \label{sec:cache}
    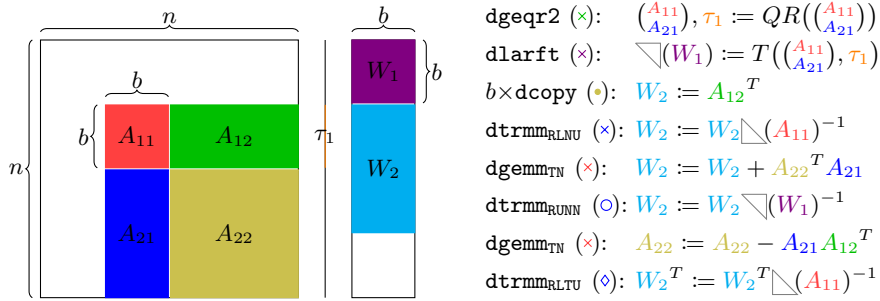
\begin{figure}[t]
    \centering

    \begin{minipage}{.49\textwidth}
        \centering
        \begin{tikzpicture}[scale=.027]
            \draw (0, 0) rectangle (127, -127);

            \filldraw[plot1] (32, -32) rectangle ++(31, -31) node[black, midway] {$A_{11}$};
            \filldraw[plot2] (64, -32) rectangle ++(63, -31) node[black, midway] {$A_{12}$};
            \filldraw[plot3] (32, -64) rectangle ++(31, -63) node[black, midway] {$A_{21}$};
            \filldraw[plot4] (64, -64) rectangle ++(63, -63) node[black, midway] {$A_{22}$};

            \draw[decorate, decoration=brace] (-4, -127) -- ++(0,   127) node[midway, anchor=east] {$n$};
            \draw[decorate, decoration=brace] (0,     4) -- ++(127,   0) node[midway, anchor=south] {$n$};
            \draw[decorate, decoration=brace] (28,  -63) -- ++(0,    31) node[midway, anchor=east] {$b$};
            \draw[decorate, decoration=brace] (32,  -28) -- ++(31,    0) node[midway, anchor=south] {$b$};
        \end{tikzpicture}
        \begin{tikzpicture}[scale=.027]
            \draw (0, 0) rectangle (0, -127);

            \filldraw[plot7] (0, -32) rectangle ++(0, -31) node[black, midway] {$\tau_1$};
        \end{tikzpicture}
        \begin{tikzpicture}[scale=.027]
            \draw (0, 0) rectangle (31, -127);

            \filldraw[plot5] (0, 0) rectangle ++(31, -31) node[black, midway] {$W_1$};
            \filldraw[plot6] (0, -32) rectangle ++(31, -63) node[black, midway] {$W_2$};

            \draw[decorate, decoration=brace] (35, 0) -- ++(0,  -31) node[midway, anchor=west] {$b$};
            \draw[decorate, decoration=brace] (0,  4) -- ++(31,   0) node[midway, anchor=south] {$b$};
        \end{tikzpicture}
    \end{minipage}
    \begin{minipage}{.49\textwidth}
        \centering
        \newcommand{\Aoo}{\textcolor{plot1}{A_{11}}}
        \newcommand{\Aot}{\textcolor{plot2}{A_{12}}}
        \newcommand{\Ato}{\textcolor{plot3}{A_{21}}}
        \newcommand{\Att}{\textcolor{plot4}{A_{22}}}
        \newcommand{\tauo}{\textcolor{plot7}{\tau_1}}
        \newcommand{\Wo}{\textcolor{plot5}{W_1}}
        \newcommand{\Wt}{\textcolor{plot6}{W_2}}
        \newcommand{\tril}[1]{{\matL{.33}{}}\!(#1)}
        \newcommand{\triu}[1]{{\matU{.33}{}}(#1)}

        \renewcommand{\arraystretch}{1.3}
        \begin{tabular}{ll}
            \refdgeqr:          &$\binom{\Aoo}{\Ato}, \tauo \coloneqq QR\bigl(\binom{\Aoo}{\Ato}\bigr)$\\
            \refdlarft:         &$\triu{\Wo}                \coloneqq T\bigl(\binom{\Aoo}{\Ato}, \tauo\bigr)$\\
            $b\times$\refdcopy: &$\Wt                       \coloneqq \Aot^T$\\
            \refdtrmmRLNU:      &$\Wt                       \coloneqq \Wt \tril{\Aoo}^{-1}$\\
            \refdgemmTN:        &$\Wt                       \coloneqq \Wt + \Att^T \Ato$\\
            \refdtrmmRUNN:      &$\Wt                       \coloneqq \Wt \triu{\Wo}^{-1}$\\
            \refdgemmTN:        &$\Att                      \coloneqq \Att - \Ato \Aot^T$\\
            \refdtrmmRLTU:      &$\Wt^T                     \coloneqq \Wt^T \tril{\Aoo}^{-1}$\\
        \end{tabular}
    \end{minipage}

    \caption{
        Memory accesses by the kernels within one step of the blocked algorithm
        \dgeqrf.  The three shapes on the left represent \dgeqrf's operands
        $A$, $\tau$, and $W$.
    }
    
    \label{fig:memacc}
    \tikzset{external/export=false}
\end{figure}

In order to predict the state of the cache throughout the execution of \dgeqrf,
we consider which parts of its operands are accessed by its kernel invocations.
\dgeqrf itself receives three operands: the input matrix $A \in \mathbb R^{n
\times n}$, an output vector $\tau \in \mathbb R^n$, and auxiliary work space $W
\in \mathbb R^{n \times b}$.  \autoref{fig:memacc} shows where within these
three memory regions the operands of the kernels invoked in one step of
\dgeqrf's blocked algorithm lie.  Since we do not consider details of the kernel
implementations, we do not make any assumptions on the patterns in which the
kernels access their operands.

For the assumed fully associative LRU cache replacement policy, identifying if a
memory region is available in cache reduces to the task of counting how many
other data elements were accessed since its last use.  To determine this count
(henceforth referred to as {\em access distance}), we scan the sequence of
kernel invocations and keep a history of the memory regions they
access\footnote{
    The length of the list can safely be restricted to to the number of kernel
    calls per iteration of the blocked algorithm.
}.  We consider the cache line as the smallest accessible memory unit:  An
access to a single data element means an access to the entire surrounding cache
line.  For each operand of a kernel invocation, we go backward through the
access history until (and including) we find its last access; thereby summing
the sizes of the accessed memory regions yields the operand's access distance.
(If the access history does not reveal a previous access, the access distance is
set to $\infty$.)

\begin{figure}[t]
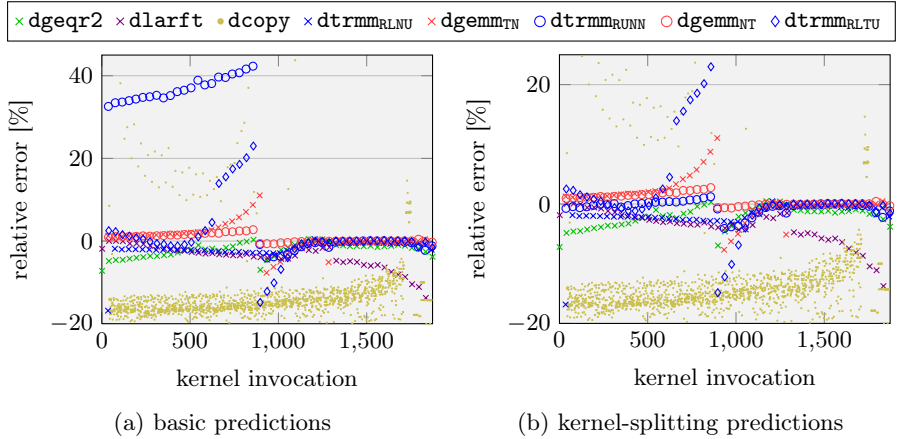

    \centering
    \ref*{fig:routinelegend}

    \vspace{1ex}
    \tikzset{external/export=true}

    \begin{subfigure}[t]{.49\textwidth}
        \centering
        \addkernelplotpic[ymin=-.2, ymax=.45]{weighted-i}
        \caption{
            basic predictions
        }
        \label{fig:i-weighted}
    \end{subfigure}
    \begin{subfigure}[t]{.49\textwidth}
        \centering
        \addkernelplotpic[ymin=-.2, ymax=.25]{split-i}
        \caption{
            kernel-splitting predictions
        }
        \label{fig:i-split}
    \end{subfigure}

    \caption{
        Basic and kernel-splitting predictions compared to in-algorithm timings
    }
    \tikzset{external/export=false}
\end{figure}

By comparing the obtained access distances to the cache size, we determine
whether the corresponding operand is expected in the cache or not.  Given these
expectations, we separately sum the sizes of the in-cache and out-of-cache
kernel operands.  These sums are then used to weight the runtime of the
corresponding timings to yield initial predictions of the instrumentation
timings, shown in \autoref{fig:i-weighted}.  Comparing to \autoref{fig:i-icooc},
our mechanism chooses (or weights) the in-cache and out-of-cache timings
correctly for most kernels.  However, the error is still $4.65\%$, because for
\refdtrmmRUNN out-of-cache is erroneously favored over in-cache.

The reason for this flaw is that (see \autoref{fig:memacc}) \refdtrmmRUNN is
preceded by the large \refdgemmTN:  This \dgemmTN's operands, which are
together larger than cache, are accumulated into \dtrmmRUNN's right-hand-side
operand's access distance.  However, since \dtrmmRUNN's right-hand-side happens
to be the output operand of the very matrix-times-vector-shaped \dgemmTN, it
appears to be left in cache.  We use this insight to extend our cache model with
a crucial assumption:  After a kernel, whose (input-)output operand is
significantly smaller than its input-only operands, we expect the (input-)output
operand to be in cache.  This assumption is implemented by splitting the memory
accesses of such a kernel into two parts: The first access contains the large
input-only operand(s), while the second only involves the small (input-)output
operand.  Therefore, the backward traversal of the access history will encounter
the latter separately and, in case it is the sought operand, terminates before
processing the cache-exceeding accesses.  The timing predictions from this
modifications (called {\em splitting predictions}) are shown in
\autoref{fig:i-split}.  Here, {\em all} kernels are chosen correctly from the
in-cache and out-of-cache timings.  As a result, the error is reduced to
$2.27\%$.

\begin{figure}[t]
    \centering
    \ref*{fig:routinelegend}

    \vspace{1ex}
    \tikzset{external/export=true}

    \begin{subfigure}[t]{.49\textwidth}
        \centering
        \tikzsetnextfilename{smoothing}
        \begin{tikzpicture}
            \begin{axis}[
                    xmax=1,
                    xmin=-1.5,
                    xlabel={$r$},
                    legend pos=north west,
                    samples=200,
                    ytick={-1,1},
                    yticklabels={$-1$, $+1$},
                ]
                \addplot[plot1, domain=-2:1] {-1 + 2 * (x > 0)};
                \addlegendentry{$\operatorname{sgn}(r)$}

                \addlegendimage{plot2};
                \addlegendentry{$\operatorname{tanh}(4 r)$}

                \addlegendimage{plot3};
                \addlegendentry{$\operatorname{tanh}(2 r)$}

                \addplot[plot2, domain=-2:1, dashed] {tanh(4 * x)};

                \addplot[plot3, domain=-2:1, dashed] {tanh(2 * x)};

                \addplot[plot2, domain=0:1]  {tanh(4 * x)};
                \addplot[plot3, domain=-2:0] {tanh(2 * x)};
            \end{axis}
        \end{tikzpicture}
        \caption{
            smoothing functions
        }
        \label{fig:smoothing}
    \end{subfigure}
    \begin{subfigure}[t]{.49\textwidth}
        \centering
        \addkernelplotpic[ymin=-.15, ymax=.1]{smooth-i}
        \caption{
            smoothed predictions
        }
        \label{fig:i-smooth}
    \end{subfigure}

    \caption{
        Smoothing functions and resulting predictions compared to in-algorithm
        timings
    }
    \tikzset{external/export=false}
\end{figure}
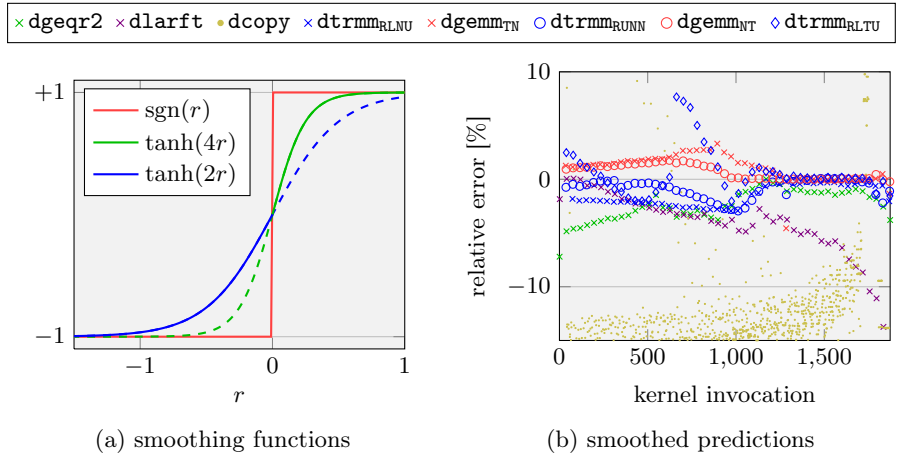

The only remaining deficiency of our predictions is in the form of severe spikes
around the transition from out-of-cache to in-cache, around the 900th kernel
invocation.  To avoid such spikes, we apply smoothing of the association of
operands with in-cache and out-of-cache.  To determine if an operator was
in-cache ($+1$) or out-of-cache ($-1$), we previously used a step function.  In
terms of the relative access distance $r = \frac{(\text{cache size}) -
(\text{access distance})}{\text{cache size}}$, this function was
$\operatorname{sgn}(r)$.  We now replace it with
$
    f(r) = \Bigl\{\begin{array}{c}
        \scriptstyle\operatorname{tanh}(\alpha r), \text{ for } r \geq 0\\
        \scriptstyle\operatorname{tanh}(\beta  r), \text{ for } r < 0
    \end{array}
$,
where $\alpha$ and $\beta$ are smoothing coefficients.  As shown in
\autoref{fig:smoothing}, $f(r)$ converges toward $\operatorname{sgn}(r)$ for
both large and small values of $r$ while showing a smooth transition of the
origin.  When applied to our predictions with empirical values of $\alpha = 4$
and $\beta = 2$, we obtain the smoothed predictions shown in
\autoref{fig:i-smooth}.  With all predictions very close to the instrumentation
timings, the error further decreases to $1.84\%$.


    \section{Results}
    \label{sec:results}
    In the previous sections we focused on a very specific setup (see
\autoref{sec:problem}). To demonstrate that our observations and models are more
broadly applicable, we now vary this setup and present the obtained accuracy
improvements of our smoothed predictions over the repeated execution timings in
\autoref{tab:compqr}.

\begin{table}[t]
    \centering
    \caption{
        Prediction improvements through cache-modeling for various scenarios.
    }
    \vspace{1ex}
    \setlength{\tabcolsep}{2pt}
    \begin{tabular}{ccccccccc}
        \toprule
                    &           &               &               &           &\ \ \  &repeated   &smoothed                  \\
        algorithm   &\#cores    &BLAS           &$n$            &$b$        &       &execution  &prediction &improvement   \\
        \midrule                
        \dgeqrf     &1          &\sc OpenBLAS   &$1{,}568$      &$32$       &       &$4.48\%$   &$1.84\%$   &$\times 2.44$ \\
        \dgeqrf     &1          &\sc OpenBLAS   &$1{,}568$      &$\bf 64$   &       &$3.15\%$   &$1.64\%$   &$\times 1.92$ \\
        \dgeqrf     &1          &\sc OpenBLAS   &$1{,}568$      &$\bf 128$  &       &$2.68\%$   &$2.13\%$   &$\times 1.26$ \\
        \dgeqrf     &1          &\sc OpenBLAS   &$\bf 2{,}080$  &$32$       &       &$5.11\%$   &$1.84\%$   &$\times 2.78$ \\
        \dgeqrf     &1          &\sc OpenBLAS   &$\bf 2{,}400$  &$32$       &       &$5.23\%$   &$1.75\%$   &$\times 2.99$ \\
        \dgeqrf     &1          &\bf ATLAS      &$1{,}568$      &$32$       &       &$3.55\%$   &$1.98\%$   &$\times 1.79$ \\
        \dgeqrf     &1          &\bf MKL        &$1{,}568$      &$32$       &       &$8.58\%$   &$4.40\%$   &$\times 1.95$ \\
        \dgeqrf     &\bf 2      &\sc OpenBLAS   &$1{,}568$      &$32$       &       &$9.58\%$   &$4.63\%$   &$\times 2.07$ \\
        \dgeqrf     &\bf 4      &\sc OpenBLAS   &$1{,}568$      &$32$       &       &$22.71\%$  &$19.75\%$  &$\times 1.15$ \\
        \bf\dtrtri  &1          &\sc OpenBLAS   &$2{,}400$      &$32$       &       &$6.70\%$   &$3.37\%$   &$\times 1.99$ \\
        \bf\dpotrf  &1          &\sc OpenBLAS   &$2{,}400$      &$32$       &       &$11.18\%$  &$7.56\%$   &$\times 1.48$ \\
        \bottomrule
    \end{tabular}
    \label{tab:compqr}
\end{table}

Although the error of our predictions remains above $1.5\%$, it is in many cases
an improvement of about a factor of $2$.  For both, increasing block-size $b$
and matrix size $n$, with a varying error for repeated executions timings, our
predictions reliably yield an error of around $2\%$.  While the picture is very
much the same, when {\sc OpenBLAS} is replaced with {\sc ATLAS}~\cite{atlas},
the error in both the repeated execution timings and our predictions increase
significantly for {\sc Intel}'s {\sc MKL}\footnote{
    For {\sc MKL}, we removed the step of splitting (input-)output from
    input-only operands in the access history; this BLAS library does not leave
    the output operand in cache.
}; however, the latter is still an improvement over the former by a factor of
$2$.  The same can be observed when doubling the number of cores to $2$.  When
we use all 4 cores of our CPU, however, the error increases drastically; this is
because every two cores share an L2 cache, while our model is designed for a
single large cache.  Finally, applying our approach also applies to other LAPACK
algorithms:  For \dtrtri (inversion of a lower triangular matrix) and \dpotrf
(Cholesky decomposition of an upper triangular matrix) it yields considerable
improvements in accuracy.


    \section{Conclusion}
    In this paper, we studied the influence of caching on the execution time of
sequences of dense linear algebra kernels within blocked algorithms.  We
established in-cache and out-of-cache timings as lower and upper bounds on the
kernel execution times within the algorithm.  We then developed a cache tracking
model that, based on a sequence of kernel invocations, predicts which memory
regions are available in cache and which are not.  With the help of this model,
we were able to combine the in-cache and out-of-cache timings into highly
accurate predictions for the actual kernel execution times.  This methodology
was shown to noticeably reduce the average error for our predictions.  The
insights and results presented in this paper constitute an important step
towards our ultimate goal of selecting and optimally configuring dense linear
algebra algorithms through performance models of the computational kernels,
without ever executing the algorithms themselves.


    \bibliographystyle{splncs}
    \bibliography{references}

\begin{thebibliography}{1}

\bibitem{modeling1}
Peise, E., Bientinesi, P.:
\newblock {Performance Modeling for Dense Linear Algebra}.
\newblock In: Proceedings of the 3rd International Workshop on Performance
  Modeling, Benchmarking and Simulation of High Performance Computer Systems
  (PMBS12). (November 2012)

\bibitem{lapacktuning}
Whaley, R.:
\newblock Empirically tuning lapack's blocking factor for increased
  performance.
\newblock In: Computer Science and Information Technology, 2008. IMCSIT 2008.
  International Multiconference on. (October 2008)  303--310

\bibitem{blocking1991}
Lam, M.D., Rothberg, E.E., Wolf, M.E.:
\newblock The cache performance and optimizations of blocked algorithms.
\newblock In: Proceedings of the Fourth International Conference on
  Architectural Support for Programming Languages and Operating Systems. ASPLOS
  IV, New York, NY, USA, ACM (1991)  63--74

\bibitem{roman}
Iakymchuk, R., Bientinesi, P.:
\newblock {Modeling Performance through Memory-Stalls}.
\newblock ACM SIGMETRICS Performance Evaluation Review \textbf{40}(2) (2012)

\bibitem{openblas}
{OpenBLAS}:
\newblock \url{http://www.openblas.net/}

\bibitem{atlas}
Whaley, R.C., Dongarra, J.:
\newblock {Automatically Tuned Linear Algebra Software}.
\newblock Technical Report UT-CS-97-366, University of Tennessee (December
  1997)

\end{thebibliography}
\end{document}